\documentclass[11pt]{article}
\usepackage{amsfonts}
\usepackage{epsfig}

\newcommand{\nwc}{\newcommand}
\nwc\eps{\varepsilon}

\nwc{\eqref}[1]{(\ref{#1})}
\nwc{\be}{\begin{equation}}
\nwc{\ee}{\end{equation}}
\nwc{\bee}{\begin{eqnarray}}
\nwc{\eee}{\end{eqnarray}}
\nwc{\ba}{\begin{array}}
\nwc{\ea}{\end{array}}

\nwc{\D}{\partial}
\nwc{\ip}[1]{{\langle #1 \rangle}}
\nwc{\ipbig}[1]{{\left\langle #1 \right\rangle}}
\nwc{\Span}[1]{\mathop{\rm span}\nolimits\{#1\}}
\nwc{\sfrac}[2]{{\textstyle\frac{ #1}{ #2}}}

\def\drawbox#1#2{
   \hbox{\vrule\vbox{\hrule width#1\vskip #2\hrule width#1}\vrule}}
\nwc{\qed}{\drawbox{.1in}{.1in}\smallskip}
\nwc{\text}{\mbox}

\nwc\bo{\mbox{Bo}}  
\nwc\calL{{\cal L}}
\nwc\calG{{\cal G}}

\newcommand\nd{\noindent}

\newcommand\IH{{\tilde{H}}}
\newcommand\IL{{\mathcal{L}}}

\newcommand\R{\mathbb{R}}
\newcommand\C{\mathbb{C}}
\newcommand\F{{\mathcal{E}}}
\newcommand\E{{\mathcal{E}_0}}
\nwc\EN{{\mathcal{E}_1}}

\newcommand\Z{\mathbb{Z}}
\newcommand\ep{\varepsilon}
\nwc{\dx}{\partial_x}
\nwc{\dy}{\partial_y}

\newcommand{\diag}{\mathop{\rm diag}\nolimits}

\newtheorem{Def}{Definition}[section]
\newtheorem{Thm}{Theorem}[section]

\newtheorem{Lem}[Thm]{Lemma}

\newtheorem{Prop}[Thm]{Proposition}

\nwc{\proof}{\noindent{\bf Proof.} }

\nwc{\CIH}[1]{\C \tilde H^{#1}}
\nwc{\SH}{S_{H}}
\nwc{\jj}{{m}}
\nwc{\US}{U^{\#}}
\nwc{\GS}{G^{\#}}

\nwc{\Ronetwo}{{\cal R}} 
\nwc{\Ronetwoinv}{{\cal P}} 
\nwc{\wtoU}{\psi}
\nwc{\HHone}{{\IH^1\times\IH^1}}
\nwc{\HHtwo}{{\IH^2\times\IH^2}}
\nwc{\vonetwo}{\tilde{V}}
\nwc{\ponetwo}{\tilde{\phi}}
\nwc{\Aonetwo}{{\cal A}}

\nwc{\phd}{\phi_\delta}
\nwc{\md}{{\cal M}_\delta}
\nwc{\fdel}{f_\delta}
\nwc{\Fd}{F_\delta}
\nwc{\Fdd}{\tilde F_\delta}
\nwc{\Forall}{\mbox{ for all\ }}
\nwc{\Sc}{S_0}
\nwc{\Sh}{S_1}
\nwc{\Mc}{M_0(\beta)}
\nwc{\Mh}{M_1(\beta)}
\nwc{\Mt}{\tilde{M}(\beta)}
\nwc{\umap}{\tilde{u}}
\nwc{\psd}{\phd}
\nwc{\CM}{\widetilde{\cal M}_\delta}
\nwc{\Lip}{{\rm Lip}}
\nwc{\pibd}{c_\pi}
\nwc{\fix}{T}
\nwc{\tfix}{\tilde{T}}
\begin{document}

\title{A host of traveling waves in a model of three-dimensional
water-wave dynamics}


\author{Robert L.  Pego\textsuperscript{1} \and  
Jos\'e Ra\'ul Quintero\textsuperscript{2}}
\date{October 2001}
\maketitle

\abstract{We describe traveling waves in a basic model for three-dimensional
water-wave dynamics in the weakly nonlinear long-wave regime.
Small solutions that are periodic in the direction of translation 
(or orthogonal to it) form an infinite-dimensional family.
We characterize these solutions through spatial dynamics, 
by reducing a linearly ill-posed mixed-type initial-value problem 
to a center manifold of infinite dimension and codimension.
A unique global solution exists for arbitrary small initial data for 
the two-component bottom velocity, specified along a single line 
in the direction of translation (or orthogonal to it).
A dispersive, nonlocal, nonlinear wave equation
governs the spatial evolution of bottom velocity.}

\medskip
\noindent {\it 2000 Mathematics Subject Classification:}
{\small 76B15, 35Q35, 35M10.}

\smallskip\noindent
{\it Abbreviated title:} 
{\small Traveling waves in a model of water-wave dynamics}
\footnotetext[1]{%
Department of Mathematics \& 
Institute for Physical Science and Technology,
University of Maryland, College Park, MD 20742 USA}
\footnotetext[2]{%
{Departamento de Matem\'aticas},
{Universidad del Valle},
{A.\ A.\ 25360, Cali, Colombia.} }


\section{\bf Introduction}

To describe steadily propagating nonlinear gravity waves on the free
surface of a three-dimensional ideal fluid, it seems natural to seek
shapes with simple symmetry, e.g., doubly periodic, or localized, say.
Hammack {\it et al.} \cite{HSS89, HMSS95} have expressed a belief that
periodic water waves may tend to form hexagonal patterns. They have produced
such waves experimentally, and have described such patterns using
theta-function solutions of the Kadomtsev-Petviashvili equation, which models
long waves of moderate amplitude propagating mainly in one direction with weak
transverse variation.  As Craig and Nicholls \cite{CrNi00} have recently
pointed out, however, in the exact water wave equations, the question of
existence of doubly periodic gravity waves exhibits the problem of small
divisors, and remains open.

In this paper we aim to develop an approach to describing steady water waves
through spatial dynamics, relaxing the assumption of periodicity in
two directions. We consider a basic isotropic model for waves in
shallow water that shares some crucial features with the exact water
wave equations. For this model we are able to describe all small waves
that translate steadily with supercritical speed and
that are periodic in the direction of translation (or orthogonal to it). 
There is in fact a host of such waves --- they form an infinite-dimensional
family. The family may be parametrized by the bottom-velocity profile
along any single line in the direction of translation (or orthogonal
to it). Fixing a supercritical wave speed 
(meaning, in dimensional terms, a speed greater than $\sqrt{gh}$, where 
$g$ is the acceleration of gravity and $h$ the undisturbed fluid
depth), for an arbitrary 
bottom-velocity profile small in a suitable Sobolev space, there
corresponds a unique globally defined traveling wave.

Simpler models of water waves exhibit the same phenomenon, 
as was discussed in \cite{HP}. 
The simplest is the linear wave equation for the surface elevation $\eta$,
given in nondimensional form by
$
\eta_{tt} = \eta_{xx}+\eta_{yy}.
$
One can find many 
traveling-wave solutions $\eta=f(x-ct,y)$ translating
with supercritical speed $|c|>1$, by solving the wave equation
$f_{yy}=(c^2-1)f_{\xi\xi}$ with 
given arbitrary initial data for the wave slope $(\eta_x,\eta_y)$ 
along the single line $y=0$ for example. 
The fact that $|c|>1$ does not violate Huyghens' principle;
these solutions are simply superpositions of two
``scissoring'' one-dimensional wave trains
that propagate with unit speed in directions oblique to the $x$-axis.

Two other simple models considered in \cite{HP} were:
(i) the exact linearized water wave equations for an inviscid
irrotational fluid without surface tension; and (ii) the 
Kadomtsev-Petviashvili equation in the KP-II case.
In each case, arbitrary small data for wave slope along a line
determine a unique traveling wave with given supercritical speed.
For the exact linearized equations, these solutions can be regarded
as superpositions of a continuum of plane waves that propagate
obliquely to the $x$-axis but translate along it with the same speed
$c$. What is intriguing
about the case of the KP equation is that one expects its
nonlinearity to disrupt the delicate superposition principle 
that apparently permits such a large family of steadily traveling
waves to exist. Since the KP equation is integrable, however,
one could speculate that the persistence of this large family of waves is due 
to some nonlinear superposition principle special to integrable systems.

In this paper, we will demonstrate that the infinite-dimensional
nature of the family of steadily propagating water wave patterns is
really a robust phenomenon, for a class of nonlinear model equations
such as were derived by Benney and Luke \cite{BL} to describe the
oblique interaction of water waves at high angles of incidence.

\section{Traveling waves for Benney-Luke equations}

The Benney-Luke equations that we consider have the form
\begin{equation}
\Phi_{tt} - \Delta \Phi + \mu (a\Delta^2 \Phi -
b \Delta\Phi_{tt}) + \eps (\Phi_t \Delta\Phi + (\nabla \Phi)^{2}_{\,t})
= 0. \label{bl-eq}
\end{equation} 
The variable $\Phi(x,y,t)$ is the 
nondimensional velocity potential on the bottom fluid boundary,
and $\mu$ and $\eps$ are small parameters:
$\sqrt{\mu}=h/L$ is the ratio of undisturbed fluid depth to typical 
wave length, and $\eps$ is the ratio of typical wave amplitude to fluid depth.
Eq.~\eqref{bl-eq} was derived for water waves without surface tension 
in \cite{BL} with $\mu=\eps$, $a=\frac16$ and $b=\frac12$.
As discussed in \cite{QP1}, Eq.~\eqref{bl-eq}
remains a formally valid water wave model in the presence of surface tension
provided $a-b=\bo-\frac13$
where $\bo$ is the Bond number (also see \cite{M-K96}). 
We take $a$ and $b$ to be positive for linear well-posedness. 
The surface elevation is related to $\Phi$ by $\eta=-\Phi_t+O(\mu,\eps)$
to leading order.

For a traveling wave solution $\Phi=u(x-ct,y)$ of \eqref{bl-eq} the wave
profile $u$ should satisfy
\be
(c^2-1)u_{xx}-u_{yy}-\mu bc^2\Delta u_{xx}+\mu a\Delta^2u
-\eps c(u_x\Delta u+|\nabla u|^2_{\,x}) = 0.
\label{bl-tw}
\ee
When the wave speed satisfies $0<c^2<\min(1,a/b)$ the traveling-wave
equation \eqref{bl-tw} is of elliptic type.
In this regime there exist finite-energy solitary waves or ``lumps,'' 
as we proved in \cite{QP1} using a variational method. 

Here we will study the complementary case $c^2>\max(1,a/b)$,
meaning simply $c^2>1$ if $\bo<\frac13$, which is physically the more
interesting case. 
In this regime the traveling-wave equation \eqref{bl-tw} has mixed type.
Consider the linear dispersion relation for \eqref{bl-tw}
for solutions of the form $u(x,y)=\exp(ikx+ily)$ with $\eps=0$:
\be
-c^2k^2+(k^2+l^2)(1-\mu bc^2k^2)+\mu a(k^2+l^2)^2 = 0.
\label{1-kl}
\ee
Solving the quadratic for $l^2$ yields
\be
l^2 =\pm \sqrt{q(k)+p(k)^2} -p(k) 
\label{1-lsq}
\ee
where
\be
p(k)= \frac12\left( \frac{1}{\mu a}+ \frac{2a-bc^2}{a}k^2\right), \qquad
q(k) = \left(\frac{bc^2-a}{a}\right)k^4+ \left(\frac{c^2-1}{\mu a}\right)k^2.
\label{1-pq}
\ee
Since $bc^2-a>0$ and $c^2-1>0$, 
for all real $k$ this yields four frequencies $l$ with
exactly two real and two purely imaginary. (See figure 1.)
Moreover, $|l|\to\infty$  as $|k|\to\infty$ in an asymptotically linear fashion.
The existence of unbounded imaginary branches of the linear
dispersion relation indicates that the initial-value problem
for \eqref{bl-tw} considered as an evolution equation in $y$
is linearly {\it ill-posed}. 

\nwc{\xsz}{3.5in}
\nwc{\ysz}{2.5in}
\nwc{\xszz}{3.5in}

\begin{figure}
\caption{Schematic plot of the branches of the dispersion relation in
\eqref{1-lsq}}
\label{Disprel}
\begin{center} \leavevmode { \hbox{
        \epsfxsize=\xsz
        \epsfysize=\ysz
        \epsffile{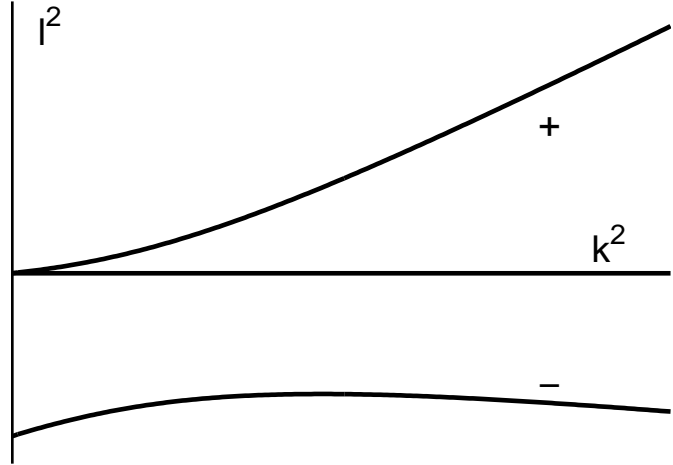}
} } \end{center}
\end{figure}

However, there is a spectral gap.
It is evident from \eqref{1-lsq} with the minus sign 
that the imaginary branches of the dispersion relation are bounded
away from the real axis --- on these branches $l^2$ is strictly negative
and $|\Im l|\ge c_0>0$ independent of $k$.
Linear modes can grow or decay or remain neutral, but modes that
grow or decay do so with rates bounded away from zero.
This linear gap structure suggests that an invariant center manifold will
exist in the nonlinear problem \eqref{bl-tw} with $\eps>0$. 
On such a manifold, modes that grow or decay in $y$ will be
slaved to the neutral modes in the infinite-dimensional space 
corresponding to real branches of the dispersion relation.

The main result of this paper involves proving that such an invariant
center manifold of infinite dimension and codimension exists and
contains all globally defined small-amplitude solutions of \eqref{bl-tw}
that are periodic in the direction of propagation.  These solutions are
determined by suitable initial data along the line $y=0$. In particular,
we show that a complete traveling wave solution of \eqref{bl-eq} is
determined uniquely by arbitrarily specifying along $y=0$ the horizontal
velocity $(\Phi_x,\Phi_y)$ at the fluid bottom, provided these data are
sufficiently small in an appropriate norm (the $H^1$ Sobolev norm).
The bottom-velocity profile evolves in $y$ according to a dispersive,
nonlinear, nonlocal wave equation obtained by restriction to the
invariant center manifold.

We obtain analogous results if we consider $x$ as the time-like variable
instead of $y$ and specify initial data along $x=0$,
and we will only sketch the analysis in this case.
The traveling wave equation \eqref{bl-tw} is not isotropic,
however, and it does not seem to be feasible in general to consider
spatial dynamics in directions other than parallel and perpendicular
to the direction of propagation of the wave.

Our bottom-velocity characterization of traveling waves for \eqref{bl-eq}
is very similar to the wave-slope characterizations 
of traveling waves for the KP-II equation and other models 
that were discussed in \cite{HP}.
For long waves of small amplitude, the variables
involved are equivalent --- to leading order, the wave slope is
directly proportional to the time derivative of bottom velocity.

It would be desirable to describe the structure of the traveling wave
solutions we find in some precise way as nonlinear superpositions of
obliquely propagating waves --- the waves exist with arbitrarily large
translation velocity $c$, reminiscent of the ``scissoring'' behavior of
superposed wave trains in the linear wave equation.  Here we do not
describe the wave structure with any accuracy beyond parametrization and
linear approximation.  But a more detailed study of the nonlinear
equation governing evolution on the center manifold may reveal further
structural information.  Perhaps doubly periodic solutions can be found
by finding $y$-periodic solutions of the governing nonlinear nonlocal
wave equation on the center manifold, for example.  We do not expect the
waves to be doubly periodic in general, however.

The ill-posed mixed-type linear structure for the Benney-Luke equations
considered here is like that for the exact water wave equations without
surface tension \cite{HP}. 
In this respect the Benney-Luke equations are a faithful model 
for the exact water wave equations. 
The problem of finding a center manifold for these systems presents technical
difficulties that have been little addressed in the literature. Both the
center subspace (spanned by the neutral modes) and its complement (spanned
by growing and decaying slave modes) are infinite-dimensional, and in both
subspaces the evolution spectrum is unbounded. 
The structure resembles a wave equation coupled
nonlinearly to an elliptic PDE. We are aware of only one work that
concerns such a mixed-type problem, a paper of Mielke \cite{Mi92}. 
Our present problem is not of the special form that Mielke considered, but
we can use some similar techniques to treat the peculiar difficulties that
arise. In an appendix we present an abstract local theorem for center
manifolds of infinite dimension and codimension which we can apply to
obtain solutions of \eqref{bl-tw}.

Ill-posed spatial evolution equations with finite-dimensional center
manifolds arise for the exact water wave problem with surface tension and
have recently been analyzed by Groves and Mielke \cite{G,GM}. 
In the exact water wave problem without surface tension, Haragus
and Pego \cite{HP} gave a linear analysis and identified some formally
conserved quantities for spatial dynamics in the nonlinear problem. But the
technical obstacles to proving there is a center manifold 
of infinite dimension and codimension in this case remain formidable.

\setcounter{equation}{0}
\section{\bf Main Result}

First, we convert the traveling wave equation \eqref{bl-tw} into a 
first order abstract differential equation in a Hilbert space 
$H$ for which the ``time-like" variable is $y$.
Since we assume $a>0$, $b>0$ and $c^2>\max(1,a/b)$, we can define
positive numbers $s$, $r$, and $d$ via
\be \label{2.srd}
s^2 = \frac{b c^{2} -a}{a}  ,\quad 
r^2 = \frac{c^2 -1}{\mu a}, \quad
d^2 = \frac{1}{\mu a}.
\ee
Introducing the variables $U_1$, $U_2$, $U_3$ and $U_4$ by
\[
U_{1}=u_{x}, \quad U_2 = u_{y},\quad  U_3=u_{yy}, \quad U_4=u_{yyy},
\]
Eq.~\eqref{bl-tw} can be viewed as the first order 
system 
\begin{equation}
\frac{dU}{dy} = A \, U + f(U) \label{eq:stw}
\end{equation}
where
\[
A=
\pmatrix{ 0 & \dx & 0 & 0 \cr
                0 & 0 & 1 & 0  \cr
                0 & 0 & 0 & 1 \cr
                s^2 \dx^3-r^2\dx & 0 & (s^2-1)\dx^2+d^2 & 0 }
, \quad U =
\pmatrix{ U_1 \cr U_2 \cr U_3 \cr U_4 }, 
\]
\[
f(U) =
\pmatrix{ 0 \cr
0 \cr
0 \cr
\ep c d^2 \left(3U_1 \, \dx U_1 + U_1 \, U_3 + 2 U_2 \, \dx U_2\right)
} .
\]

We will restrict ourselves to consider only solutions periodic in
$x$ with fixed period, which we take to be $2\pi$ by rescaling.
By scaling amplitude we can assume $\ep=1$.
We shall keep the parameters $a$, $b$, $c$ and $\mu$ fixed.

Given an integer $k\ge0$, let $\IH^k$
denote the Sobolev space of $2\pi$-periodic functions on $\R$
whose weak derivatives up to order $k$ are square-integrable.
Then $\IH^k$ is a Hilbert space with norm given by
\[
\|u\|_{\IH^k}^2 = \sum_{j=0}^k \int_0^{2\pi} |\dx^j u|^2\,dx
\]
To study \eqref{eq:stw}
we introduce Hilbert spaces $H$ and $X$ defined by
\begin{eqnarray}
H&=& \IH^{1} \times \IH^{1} \times \IH^{0} \times \IH^{-1},\\
X&=& \IH^{2} \times \IH^{2} \times \IH^{1} \times \IH^{0} .
\end{eqnarray}
Note that $X$ is densely embedded in $H$.

Our main result is the following. 
\begin{Thm}(Traveling wave solutions via dynamics in $y$) \label{th:main}  
There are positive constants $\delta_1$ and $C_1$ with the following
property:
Given any initial conditions of the form
\[
(U_1(0), U_2(0))= (w_1, w_2) 
\]
in $\IH^2\times \IH^2$ 
such that 
$\left\|(w_1, w_2) \right\|_{ \IH^1\times\IH^1} \leq \delta_1$, 
Eq.~(\ref{eq:stw}) has a unique 
global classical solution $U\in C^1(\R,H)\cap C(\R, X)$
such that
$\|U(y)\|_{H} \leq C_1\delta_1$ for all $y\in\R$.
The map taking initial conditions 
to the solution is Lipschitz continuous
from $\IH^1\times\IH^1$ to $C([-T,T],H)$,
for any $T>0$.

Moreover, the first two components of the solution satisfy
a dispersive, nonlinear, nonlocal wave equation of the form
\be\label{eq:wave}
\frac{d}{dy}\pmatrix{U_1\cr U_2} =
\pmatrix{0&\dx\cr S\dx&0}\pmatrix{U_1\cr U_2}+
\pmatrix{0\cr g(U_1,U_2)},
\ee
in which the map 
$g\colon \IH^1\times\IH^1\to \IH^1$ is Lipschitz with
$g(0)=0$, $Dg(0,0)=0$, and where the nonlocal
linear operator $S$ (defined precisely in \eqref{eq:Shat}) 
yields the real linear dispersion relation
given by \eqref{1-lsq} with the plus sign.
\end{Thm}

It is evident from the stability estimates that, 
for any small initial data $(w_1,w_2)$ in $\HHone$,
\eqref{eq:stw} has a global weak solution $U\in C(\R,H)$ which satisfies the
same stability estimates.

For traveling wave solutions via dynamics in $x$, the roles of $x$ and 
$y$ are simply interchanged throughout. With
$U=(u_y,u_x,u_{xx},u_{xxx})$ one obtains a first order
system of the same form as in \eqref{eq:stw} except that
\be \label{2.srdx}
s^2 = \frac{a}{b c^{2} -a},\quad 
r^2 = \frac{1}{\mu(bc^2- a)}, \quad
d^2 = \frac{c^2 -1}{\mu (bc^2-a)}.
\ee
and $f_4(U)= -\ep cr^2(3U_2U_3+U_2\dy U_1+2U_1\dy U_2)$.
Exactly the same theorem as \ref{th:main} holds in this case, with
$y$ replaced by $x$. 

To prove these results, we will first apply the abstract local center manifold
theorem in the appendix to our problem (see section 3). 
This will show that with respect to
an invariant-subspace decomposition $H=H_0\oplus H_1$ of $H$ into
a center subspace $H_0$ and a hyperbolic subspace $H_1$,  equation
\eqref{eq:stw} has a local invariant manifold given by the graph of a 
Lipschitz function $\phd\colon H_0\to H_1\cap X$. 

Next, we will use a conserved energy functional $\F$ 
to prove that the zero solution is stable on the local center manifold,
and infer the global existence of classical solutions for small data 
(see section 4).  The energy functional is indefinite in general, but
on the center subspace and center manifold it is coercive with respect to the
$H$ norm.  

Finally, we will show in section 5 that the center subspace and center manifold
are conveniently parametrized by the first two components of $U$,
which correspond physically to the horizontal velocity on the fluid bottom. 
We will then obtain global classical solutions for (\ref{eq:stw}) with
data $(w_1, w_2)$ in the space $\IH^2\times \IH^2$ 
that are small in the $\HHone$ norm, and verify that 
solutions satisfy an equation of the form in \eqref{eq:wave}.

\setcounter{equation}{0}

\section{\bf Existence of a local center manifold}
 
The goal in this section is to verify the hypotheses of Theorem
\ref{T:1} in the appendix (an abstract center manifold theorem) to obtain
existence of a local center manifold for the system \eqref{eq:stw}.
These hypotheses come in three groups: (1) basic structural conditions
on $A$ and $f$, (2) existence of a spectral decomposition of 
$H$ as a direct sum of a center subspace $H_0$ and a hyperbolic
subspace $H_1$, with a corresponding decomposition of $X$, and
(3) solvability conditions on linear equations of evolution 
in each subspace.

\subsection{Structural conditions}

First consider the basic structural conditions. It is straightforward
to check that $A\in\IL(X,H)$,
the space of bounded linear operators from $X$ to $H$.
The map $f$ is bilinear, and with $ U, V \in H$ it is easy to check that
\be
\|f(U+V) -f(U)\|_{X} \le (\|U\|_H+\|V\|_H)\|V\|_H.
\label{fest}
\ee
By consequence, $f\colon H\to X$ is smooth, and clearly $f(0)=0=Df(0)$.
It is an interesting feature of this problem that the nonlinear map
$f$ exhibits a gain of regularity. We will exploit this feature
to help find classical solutions.

\subsection{Spectral decomposition}

We will construct the desired spectral decompositions
of $H$ and $X$ by using a complete set of eigenfunctions found
via Fourier transform. 

Any element $U\in H$ or $X$ can be represented by a Fourier series
\be
U= \sum_{k \in \Z}
\hat{U}(k)e^{ikx}.
\label{3-uhat}
\ee
In terms of the vector Fourier coefficients
the norms in $H$ and $X$ may be given by 
\be\label{FTnorm}
\|U\|^{2}_{H} = \sum_{k \in \Z}|\SH(k) \hat{U}(k)|^2,
\qquad 
\|U\|^{2}_{X} = \sum_{k \in \Z}(1+k^2)|\SH(k) \hat{U}(k)|^2,
\ee
where
\[
\SH(k)= \diag \left\{(1+k^2)^{1/2}, (1+k^2)^{1/2}, 1,
(1+k^2)^{-1/2}\right\}.
\]
For $U\in X$ we have $\widehat{AU}(k)=\hat{A}(k)\hat{U}(k)$ where
\[
\hat{A}(k)= \pmatrix{ 0 &     ik    &    0    &     0 \cr 
                     0   & 0   &    1    &     0  \cr 
                     0   &     0     & 0 & 1 \cr 
                -i(s^2 k^3+r^2k) & 0 & (1-s^2)k^2+d^2 & 0
}.
\]

If $\lambda\in\C$ is an eigenvalue with eigenfunction $U$, it
must be that $(\hat{A}(k)-\lambda I)\hat{U}(k)=0$ for all $k$,
so that for some $k$, $\hat{U}(k)$ is an eigenvector of $\hat{A}(k)$
with eigenvalue $\lambda$. Let us write 
\be\label{Qdef}
Q(\lambda, k)= \det(\hat{A}(k)-\lambda I)=
\lambda^4 - 2p(k)\lambda^2 -q(k)
\ee
where $p(k)$ and $q(k)$ are given in \eqref{1-pq}.
An eigenvalue $\lambda$ must satisfy $Q(\lambda,k)=0$ for some $k$.
For each nonzero integer $k$ we find four distinct eigenvalues,
two purely imaginary and two real, which we denote as follows:
\be\label{eigs}
\begin{array}{ll}
\lambda_1(k) = 
i\sqrt{-p(k)+\sqrt{p(k)^2+q(k)}}, &
\lambda_2(k)= -\lambda_1(k),\\[3pt]
\lambda_3(k) = 
\sqrt{p(k)+\sqrt{p(k)^2+q(k)}}, &
\lambda_4(k)= -\lambda_3(k),
\end{array}
\ee
For $k=0$ the eigenvalue $\lambda_1(0)=\lambda_2(0)=0$ is a double zero of 
$Q(\lambda,0)$, and $\lambda_3(0)=-\lambda_4(0)=d$.
As $k\to\infty$ we have that $\lambda_1(k)\sim isk$, $\lambda_3(k)\sim k$.
The sign of $1-s^2$ is not determined, but $\lambda_3(k)\ge\alpha>0$
for some constant $\alpha$ independent of $k$.
For each nonzero eigenvalue $\lambda$ note that 
\[
\D_\lambda Q(\lambda, k)= 4\lambda^3-4p(k)\lambda =
\frac{2}{\lambda}\left(\lambda^4+q(k)\right) 
\not = 0 .
\]

Corresponding to each nonzero eigenvalue $\lambda_\jj(k)$,
the matrix $\hat{A}(k)$ has right eigenvector $v_\jj(k)$ and
left eigenvector $w_\jj(k)$ given by
\begin{eqnarray} 
v_\jj(k) &=& \left(ik, \lambda_\jj(k), 
\lambda_\jj(k)^2, \lambda_\jj(k)^3\right)^T, \label{revec}\\
w_\jj(k) &=& 
\left( \frac{q(k)}{ik
\lambda_\jj(k)}, \frac{q(k)}{\lambda_\jj(k)^2}, 
\lambda_\jj(k), 1 \right)
\cdot\frac{1}{\D_\lambda Q(\lambda_\jj(k),k)}
.\label{levec}
\end{eqnarray}
For the zero eigenvalue with $k=0$ there is a two-dimensional
eigenspace and we let
\be
\begin{array}{ll}
v_1(0)=\left(1,0,0,0\right)^T, &
v_2(0)=\left(0,1,0,0\right)^T, \\
w_1(0)=\left(1,0,0,0\right), &
w_2(0)=\left(0,1,0,-d^{-2}\right).
\end{array}
\ee
The eigenvectors are normalized so that $w_\jj(k) \cdot v_\jj(k) = 1$.
Introducing the matrices
\be
W(k)=\pmatrix{w_1(k)\cr w_2(k)\cr w_3(k)\cr w_4(k)},
\quad
V(k)=\pmatrix{v_1(k),v_2(k),v_3(k),v_4(k)},
\label{3-wv}
\ee
we have $W(k)\cdot V(k)=I$ and 
$W(k)\hat{A}(k)V(k)=
\diag\{ \lambda_1(k), \lambda_2(k), \lambda_3(k), \lambda_4(k)\}$
for all $k$.

Given an element $U$ in $H$ may write 
\be
\hat{U}(k)=V(k)\US(k) \qquad
\mbox{where\ \ }
\US(k)=\pmatrix{\US_1(k)\cr \US_2(k)\cr \US_3(k)\cr \US_4(k)}=W(k)\cdot\hat{U}(k),
\label{3-us}
\ee
and hence we have the representations
\be
U = 
\sum_{k\in\Z}\sum_{\jj=1}^4 e^{ikx} v_\jj(k)\US_\jj(k),
\qquad
AU=
\sum_{k\in\Z}\sum_{\jj=1}^4 e^{ikx} v_\jj(k)\US_\jj(k)\lambda_\jj(k).
\label{3-urep}
\ee
Because $|\lambda_\jj(k)|$ grows asymptotically linearly in $k$,
it is not difficult to check that
in terms of the coefficient vectors $\US(k)$ we have the following
equivalences of norms:
\be\label{normeq}
\|U\|_H^2 \sim \sum_{k\in\Z} (1+k^2)^2|\US(k)|^2,
\qquad
\|U\|_X^2 \sim \sum_{k\in\Z} (1+k^2)^3|\US(k)|^2.
\ee

We define the projections $\pi_0$ and $\pi_1$ by 
\be\label{pidef}
\pi_0U = 
\sum_{k\in\Z}\sum_{\jj=1}^2 e^{ikx} v_\jj(k)\US_\jj(k),
\qquad
\pi_1U = 
\sum_{k\in\Z}\sum_{\jj=3}^4 e^{ikx} v_\jj(k)\US_\jj(k).
\ee
From the equivalences in \eqref{normeq} it is evident that 
$\pi_0$ and $\pi_1$ are bounded on $H$ and on $X$ with 
$\pi_0+\pi_1=I$, and it is clear that $AX_j\subset H_j$ 
where $X_j=\pi_jX$ and $H_j=\pi_jH$ for $j=0,1$.
This yields the spectral decompositions 
$H = H_0 \oplus H_1$ and $ X = X_0 \oplus X_1$ 
with the properties required in the appendix.

\subsection{Solvability conditions for linear dynamics.}

First we consider the center subspace $H_0$.
We define a family of linear operators $\{S_0(t)\}_{t\in\R}$ on $H_0$ by 
\be\label{S0def}
S_0(t)U= \sum_{k\in\Z}\sum_{\jj=1}^2 e^{ikx}
v_\jj(k)\US_\jj(k) e^{\lambda_\jj(k)t}.
\ee
Since for $\jj=1$ and $2$, $\lambda_\jj(k)$ is pure imaginary
and its magnitude grows asymptotically linearly in $k$, it is
straightforward to show that the family $\{S_0(t)\}_{t\in\R}$ 
is a bounded $C^0$-group on $H_0$ with infinitesimal generator
$A_0=A|_{X_0}$. This establishes that the hypothesis (H0)
of the appendix holds.

Next we seek to verify condition (H1) in the hyperbolic subspace $H_1$.
For consistency with the notation in the appendix we replace $y$ by
$t$ in the rest of this section.
We must consider the inhomogeneous linear equation 
\be\label{h1eq}
\frac{d}{dt}U(t)=A_1U(t)+G(t)
\ee
where $A_1=A|_{X_1}$. 
Recall that $\lambda_3(k)\ge\alpha>0$
for all $k$. Let $0\le\beta<\alpha$ and let $G\in C(\R,X_1)\cap H_1^\beta$,
where as in the appendix for any Banach space $Y$ we let
\be\label{Ybeta}
Y^\beta = 
\{u\in C(\R,Y)\mid \|u\|_{Y^\beta}:=\sup_t e^{-\beta|t|}\|u\|_Y<\infty\}.
\ee

Suppose $U\in C^1(\R,H_1)\cap C(\R,X_1)$ is a solution belonging to 
$H_1^\beta$.
Then applying the Fourier transform in $x$ and multiplying
by the matrix $W(k)$ yields
\be\label{hseq}
\frac{d}{dt} \US_\jj(k,t)=\lambda_\jj(k)\US_\jj(k,t)+\GS_\jj(k,t)
\ee
for all $k\in\Z$, $t\in\R$ and $\jj=3,4$.
The functions $\GS_\jj(k,\cdot)$ and $\US_\jj(k,\cdot)$ belong
to $\R^\beta$ ($Y=\R$ in \eqref{Ybeta}). 
Since $|\lambda_\jj(k)|\ge\alpha>\beta$, it is necessarily the case that
\be
\US_3(k,t)= \int_\infty^t e^{\lambda_3(k)(t-\tau)} \GS_3(k,\tau)\,d\tau,
\qquad
\US_4(k,t)= \int_{-\infty}^t e^{\lambda_4(k)(t-\tau)} \GS_4(k,\tau)\,d\tau.
\label{h1u34}
\ee

Consequently any solution of \eqref{h1eq} in $H_1^\beta$ is unique.
We claim that the formulas \eqref{h1u34} together with the representation
for $U=\pi_1U$ in \eqref{pidef} yield existence of a solution in
$H^1_\beta$. 
For this purpose it will be convenient to decompose Eq.~\eqref{h1eq}
using projections into the ``unstable" and ``stable" subspaces. 
These projections are defined by 
\be\label{suproj}
\pi_uU = 
\sum_{k\in\Z}e^{ikx} v_3(k)\US_3(k),
\qquad
\pi_sU = 
\sum_{k\in\Z}e^{ikx} v_4(k)\US_4(k)
\ee
for $U\in H$. 
Clearly $\pi_u$ and $\pi_s$ are bounded on $H$ and $X$ 
and $\pi_u+\pi_s=\pi_1$. 

We next introduce a Green's function operator 
$S(t)$ defined for nonzero $t\in\R$ by
\be\label{sudef}
S(t)U = \cases{\displaystyle
-\sum_{k\in\Z}e^{ikx} v_3(k)\US_3(k)e^{\lambda_3(k)t}
& for $t<0,$ \cr\displaystyle
\phantom{-}\sum_{k\in\Z}e^{ikx} v_4(k)\US_4(k) e^{\lambda_4(k)t}
& for $t>0$.}
\ee
Due to \eqref{normeq}, $S(t)$ is equivalent to a multiplication operator
on a sequence space $\ell^2$, so it is easy to verify that for some 
constant $C$ independent of $t$ we have the following 
norm bounds:
\begin{eqnarray}
\|S(t)\|_{\IL(Y)} &\le& C e^{-\alpha|t|}
\qquad\mbox{($Y=H$ or $X$),} \label{SHbd}\\
\|S(t)\|_{\IL(H,X)} &\le& \cases{
C e^{-\alpha t}& for $\alpha|t|\ge1$,\cr
C|t|^{-1} & for $\alpha|t|\le 1$,}
\label{SHXbd}
\\
\|S(t)-\pi_s\|_{\IL(X,H)}+\|S(-t)+\pi_u\|_{\IL(X,H)}&\le& Ct
\quad\mbox{for $t>0$}.
\label{SXHbd}
\end{eqnarray}
Here $\IL(Y)$ and $\IL(H,X)$ respectively 
denote the space of bounded operators on $Y$, and from $H$ to $X$.
To get these bounds one uses the facts that $\lambda_3(k)=-\lambda_4(k)$
is bounded below by $\alpha>0$ and grows linearly in $k$, so
\[
\|\pi_1U\|_X^2\sim 
\sum_{k\in\Z} \sum_{\jj=3}^4 
(1+k^2)^2|\lambda_\jj(k)\US_\jj(k)|^2,
\]
together with the facts that for $t>0$,
\[
\sup_{\lambda\ge\alpha}\lambda^{-1}|e^{-\lambda t}-1| \le
t, \qquad
\sup_{\lambda\ge\alpha}\lambda e^{-\lambda t } = \cases{
\alpha e^{-\alpha t }& for $\alpha t \ge1$,\cr
1/e t & for $\alpha t \le1$.}
\]
Also, $S$ is $C^1$ from $\R\setminus\{0\}$ to $\IL(H)$ 
with $dS(t)/dt=A_1S(t)$, and 
$S(t)\to\pi_s$ (resp. $-\pi_u$) strongly as $t\to0^+$ (resp. $0^-$).
Therefore 
the families $\{S(t)\}_{t>0}$ and $\{-S(-t)\}_{t>0}$ are
analytic semigroups in $\pi_sH$ and $\pi_uH$ respectively
\cite[p.62]{Pa}.

Eqs.~\eqref{h1u34} yield the formula 
\be\label{h1u}
U(t)=\int_{-\infty}^\infty S(t-\tau)G(\tau)\,d\tau
\ee
for the solution of \eqref{h1eq}. We wish to show that 
$U\in C(\R,X)$, $U\in H_1^\beta$, and $dU/dt$ exists in $H$
and satisfies \eqref{h1eq}. For the first step we note that
\[
U(t)= \int_{|s|\le\alpha^{-1}} S(s)G(t-s)\,ds+ 
      \int_{|s|\ge\alpha^{-1}} S(s)G(t-s)\,ds.
\]
Using \eqref{SHbd} and $G\in C(\R,X)$ it is clear that the
first term is in $C(\R,X)$. For the second term we use \eqref{SHXbd}
and $G\in H^\beta$ to see that the integral converges in $X$ uniformly
on compact sets in $t$.

It follows that $U\in C(\R,X)$. Using \eqref{SHbd} we have
that for $Y=H$ or $X$, if $G\in Y^\beta$ then
\be\label{Kest}
e^{-\beta|t|}\|U(t)\|_Y \le C\|G\|_{Y^\beta}
\int_{-\infty}^\infty e^{-\alpha|s|+\beta(|t-s|-|t|)}\,ds
\le \frac{2C}{\alpha-\beta}\|G\|_{Y^\beta}.
\ee
This shows that $U\in Y^\beta$ and establishes the estimates
required in hypothesis (H1).

It remains to show $U=\pi_sU+\pi_uU$ is differentiable in $H$ 
and satisfies \eqref{h1eq}.
We check this for the two terms separately. For $h>0$ we compute
\begin{eqnarray*}
\frac{\pi_sU(t+h)-\pi_sU(t)}{h} &=&
\left(\frac{S(h)-\pi_s}{h}\right) \pi_sU(t) 
+ \frac1h\int_0^h (S(\tau)-\pi_s)G(t+h-\tau)\,d\tau\\&&  
+\ \frac1h\int_t^{t+h} \pi_sG(\tau)\,d\tau.
\end{eqnarray*}
Using \eqref{SXHbd} and $G\in C(\R,X)$, as $h\to0^+$ we deduce 
that the 
next-to-last term converges to zero in $H$ and the last term converges
to $\pi_sG(t)$. Moreover the first term converges to $A_1\pi_sU(t)$.
Hence the right derivative exists and satisfies
$D_+\pi_sU(t)=A_1\pi_sU(t)+\pi_sG(t)$, so is continuous into $H$.
It follows that $\pi_sU$ is differentiable. We may treat $\pi_uU$
in a similar way, and conclude that $U$ is differentiable and satisfies
\eqref{h1eq}.

This completes the verification of hypothesis (H1) in the appendix.
Since all the hypotheses of Theorem~\ref{T:1} have been verified, 
we have established that system \eqref{eq:stw} admits a local center manifold
having the properties stated in the Theorem.

\setcounter{equation}{0}
\nwc{\usp}{{\US_1(k)+\US_2(k)}}
\nwc{\usm}{{\US_1(k)-\US_2(k)}}

\section{\bf Global existence and stability }
Our aim in this section is to establish the global existence
of classical solutions on the local center manifold, for initial
data that is small in $H$-norm. This we do by establishing
that the zero solution is stable on the center manifold, which
is given by the graph of a function $\phd\colon H_0\to X_1$.
We shall exploit an energy functional that is conserved in ``time''
for classical solutions. We shall prove the following result.

\begin{Thm}\label{th:stab}
(Stability on the center manifold)
Let $\phd$ be given by applying Theorem~\ref{T:1} to \eqref{eq:stw}.
There exist positive constants $\delta_2$ and $C_2$ such that,
for any $\xi\in X_0$ with $\|\xi\|_H\le \delta_2$, 
there is a unique classical solution $U$ on $\R$ to \eqref{eq:stw} such that
$\pi_0 U(0)=\xi$ and 
$\|U(y)\|_H\le 2C_2\|\xi\|_H$ for all $y\in\R$. 
Moreover, for any $T>0$ the map taking $\xi$ to $U$ is Lipschitz continuous
from $H_0$ to $C([-T,T],H)$.
\end{Thm}

We define the functional $\F\colon H \to {\R}$ by
\be\label{4-E}
\F(U) = \E(U)+\EN(U)
\ee
where the quadratic part is
\begin{eqnarray}
\E(U)&:=& \frac{1}{2\pi} \int_{0}^{2 \pi}
\biggl(
r^2|U_1|^2 + s^2 |\dx U_1|^2 + d^2|U_2|^2 - (s^2-1)|\dx U_2|^2 + |U_3|^2
\biggr) \,dx \nonumber\\[3pt]
&&-\ \frac{1}{\pi}
\left(U_4,U_2\right)_{-1,1}, 
\label{4-e0}
\end{eqnarray}
and the remaining part is 
\be\label{4-en}
\EN(U) = 
-\frac{\ep cd^2}{2\pi}
\int_0^{2\pi}  (U_1^3 - U_1U_2^2) \,dx .
\ee
Above, $\left( \cdot, \cdot
\right)_{-1,1}$ denotes the natural pairing between $\IH^{-1}$ and
$\IH^{1}$. Clearly $\F$ is smooth from $H$ to $\R$.
If $U \in C^1({\R},H)$ is a classical solution of the first order 
equation \eqref{eq:stw}, then for all $y\in\R$ we find
\[
\displaystyle{\frac{d}{dy}} \F(U(y)) = 0.
\]
In other words, $\F$ is conserved along classical solutions of 
~(\ref{eq:stw}).

In the case of dynamics in $x$, when $x$ and $y$ are interchanged,
the quadratic part of the energy has the same form and the remaining part
is replaced by 
\be\label{4-en-x}
\EN(U) = 
\frac{-\ep cr^2}{\pi}
\int_0^{2\pi}  U_2^3 \,dy .
\ee

Note that neither $\F$ nor $\E$ is necessarily positive. However, we will 
establish that their restrictions to the center manifold are 
positive. We will first show that $\E$ is positive in the center space 
$H_0$. 
\begin{Lem} \label{lm:E0}
There is a positive constant $C_0$ such that for any $U \in H_0$, 
\[
\left(1/C_0 \right) \|U\|^2_{H} \leq \E(U) \leq C_0 \|U\|^2_{H}.
\]
\end{Lem}
\nd {\bf Proof.}
Using the Fourier series representation \eqref{3-uhat}
we find
\be
\E(U) = \sum_{k\in\Z} \E(\hat U(k)e^{ikx}).
\label{4-erep}
\ee
Now
\be\label{4-uhat}
\hat U(0)=\pmatrix{\US_1(0)\cr \US_2(0)\cr 0\cr 0}, \quad
\hat U(k)=\pmatrix{ik (\usp) \cr \lambda_1(k)(\usm)\cr
\lambda_1(k)^2(\usp)\cr \lambda_1(k)^3(\usm)},
\ee
so 
\[
\E(\hat U(0)) = r^2|\US_1(0)|^2+d^2|\US_2(0)|^2,
\]
and since $\lambda_1(k)$ is purely imaginary we compute
for $k\ne0$ that
\begin{eqnarray}
\E(\hat U(k)e^{ikx})&=&
(r^2+s^2k^2)
|\hat U_1(k)|^2
+(d^2+(1-s^2)k^2)
|\hat U_2(k)|^2
+|\hat U_3(k)|^2
-2 \hat U_4(k)\overline{\hat U_2(k)} \nonumber \\
&=&
(q(k)+|\lambda_1(k)|^4) |\usp|^2 \nonumber \\
&& +\ |\lambda_1(k)|^2(2p(k)-2\lambda_1(k)^2 )|\usm|^2.
\label{4-ecomp}
\end{eqnarray}
The quantity
\be\label{4-l}
L(k):=2p(k)-2\lambda_1(k)^2 = 
2\sqrt{p(k)^2+q(k)} 
\ee
is positive, and for $|k|\ge1$, the quantity $|\lambda_1(k)|^2L(k)$
is bounded above and below by a constant times
$(1+k^2)^2$. 
The same is true for $q(k)+|\lambda_1(k)|^4$.
Therefore by \eqref{normeq} we obtain the desired equivalence:
\begin{eqnarray}
\E(U)&\sim&
\sum_{k\in\Z} (1+k^2)^2\left(|\usp|^2+|\usm|^2\right)\nonumber\\
&=&
2\sum_{k\in\Z} (1+k^2)^2\left(|\US_1(k)|^2+|\US_2(k)|^2\right) 
\sim \|U\|_H^2.
\qquad\qed
\label{4-eeq}
\end{eqnarray}

Next we control the energy on the center manifold, which 
is given by the graph of a function 
$\phd\colon H_0\to X_1$ with the properties stated in Theorem~\ref{T:1}.

\begin{Lem} \label{lm:e-cm}
Let $\phd$ be given by applying Theorem~\ref{T:1} to \eqref{eq:stw}.
Then there exist positive constants $\delta_2$ and $C_2$ such that
for all $\xi\in H_0$ with $\|\xi\|_H<\delta_2$ we have
\[
\frac{1}{C_2}\|\xi\|_H^2 \le \F(\xi+\phd(\xi)) 
\le C_2\|\xi\|_H^2.
\]
\end{Lem}

\proof
By Theorem~\ref{T:1}, $\|\phd(\xi)\|_H=o(\|\xi\|_H)$ 
as $\|\xi\|_H\to0$, 
and since
$\EN(\xi+\phd(\xi))=O(\|\xi\|_H^3)$ it is easy to see that
\begin{eqnarray}
\F(\xi+\phd(\xi))
&=&\E(\xi)+O(\|\xi\|_H\|\phd(\xi)\|_H) +\EN(\xi+\phd(\xi))\\
&=& \E(\xi)+o(\|\xi\|_H^2)
\label{4-fest}
\end{eqnarray}
as $\|\xi\|_H\to0$. The result follows using Lemma~\ref{lm:E0}.
\qed

Let us now proceed to prove Theorem \ref{th:stab}.
Let $\delta_2$ and $C_2$ be given by Lemma \ref{lm:e-cm}. 
We may assume $2C_2\delta_2<\delta$ by making $\delta_2$ smaller
if necessary.
Suppose $\xi\in X_0$ with $\|\xi\|_H\le\delta_2$. 
We invoke Theorem~\ref{T:1} and obtain existence of a continuous function 
$U$ from $\R$ into the center manifold
$\md=\{\zeta+\phd(\zeta):\zeta\in X_0\}$ such that $\pi_0U(0)=\xi$,
which is a classical solution of \eqref{eq:stw} on any 
open interval $J\subset\R$ containing $0$ such that 
$\|\pi_0U(y)\|_H<\delta$ for all $y\in J$.

By a straightforward continuation argument, to show that $U$
is a classical solution on $\R$ and satisfies the estimate
claimed in Theorem \ref{th:stab}, it suffices to establish
an appropriate a priori estimate. Namely, it is enough
to show that on any 
open interval $J\subset\R$ containing $0$ such that 
$\|\pi_0U(y)\|_H<\delta$ for all $y\in J$,
we have that
$\|\pi_0U(y)\|_H\le C_2\|\xi\|_H$ for all $y\in J$.

Since $U$ is a classical solution on such an interval, we
may use the fact that $\F(U(y))$ is constant along with
Lemma \ref{lm:e-cm} to deduce that for any $y\in J$,
\be\label{4-eest}
\frac{1}{C_2}\|\pi_0U(y)\|_H^2 \le \F(U(y)) =\F(U(0))
\le C_2\|\xi\|_H^2.
\ee
This establishes the desired a priori estimate, proving
the existence part of the Theorem. The statement regarding
Lipschitz dependence follows from Proposition \ref{R:2.4}.

To prove the uniqueness statement, suppose 
$U$ is a classical solution on $\R$ to \eqref{eq:stw} such that
$\pi_0 U(0)=\xi$ and 
$\|U(y)\|_H\le 2C_2\|\xi\|_H$ for all $y\in\R$. 
Since $2C_2\|\xi\|_H<\delta$,
from Theorem~\ref{T:1}(v) it follows $U(y)$ must lie on the center manifold
$\md$ for all $y\in\R$.
Then $U$ is determined by $U_0=\pi_0U$, which by Theorem~\ref{T:1}(iii)
is the unique classical solution of the equation 
\be
\label{eq:U0}
 \frac{d}{dy}U_0(y) = A_0U_0(y) + \pi_0 f(U_0(y)+\phd(U_0(y)))
\ee
with $U_0(0)=\xi$.

\setcounter{equation}{0}

\section{\bf Parametrization and evolution on the center manifold}

The last steps needed to prove Theorem \ref{th:main} involve showing
that global solutions $U$ on the center manifold can be characterized
by their initial data in just the first two components. 
In order to accomplish this characterization, by
Theorem \ref{th:stab} it will suffice to establish
a suitable correspondence between the first two components of $U(0)$
and the center-subspace projection $\pi_0U(0)$.
In this section, we denote the restriction of any 
$U=(U_1,U_2,U_3,U_4)^T$ on the first two components by 
\be\label{5-r12}
\Ronetwo U = (U_1,U_2).
\ee

\begin{Thm} \label{th:init}
Let $\phd$ be given by applying Theorem~\ref{T:1} to \eqref{eq:stw}.
There exist positive constants $\delta_3$ and $C_3$ with the
following property. For any $w=(w_1,w_2)\in\HHone$ such that 
$\|w\|_{\HHone}<\delta_3$, there exists a unique
$\xi\in H_0$ such that $\|\xi\|_H\le C_3\|w\|_\HHone$ and 
\be\label{5-w}
w=\Ronetwo (\xi+\phd(\xi)).
\ee
The map $w\to\xi$ is Lipschitz continuous, and if $w\in\HHtwo$ then
$\xi\in X_0$.
\end{Thm}

To prove this, first we study the restriction $\Ronetwo$ on the center subspace.

\begin{Lem} \label{lm:r12}
The map $\Ronetwo$ yields simultaneous isomorphisms
$H_0 \cong \HHone$ and $X_0\cong\HHtwo$.
\end{Lem}

\proof 
Given any $U=\pi_0 U$ in $H_0$ or $X_0$, we use the representation
\eqref{pidef} to write
\be\label{5-ru}
\Ronetwo U = \sum_{k\in\Z} e^{ikx}
\vonetwo(k)\pmatrix{\US_1(k)\cr\US_2(k)}.
\ee
where
\be\label{5-v12}
\vonetwo(0)=\pmatrix{1&0\cr0&1},\quad
\vonetwo(k)=
\pmatrix{ik & ik\cr \lambda_1(k)&-\lambda_1(k)}
\ee
for $k\ne0$. Since $|\lambda_1(k)|$ grows asymptotically linearly in $k$,
it follows that 
\be\label{5-rest}
|\widehat{\Ronetwo U}(k)|^2 \le C(1+k^2)(|\US_1(k)|^2+|\US_2(k)|^2).
\ee
From the equivalences \eqref{normeq} it follows that
$\Ronetwo$ is bounded from $H_0$ to $\HHone$ and from $X_0$ to $\HHtwo$.
Moreover, $\Ronetwo$ is one-to-one, since if $\Ronetwo U=0$ then
$\widehat{U}(k)=0$ for all $k$ since $\vonetwo(k)$ is invertible.

To see that $\Ronetwo$ yields an isomorphism, we observe that its inverse
is given by the prolongation formula 
$U=\Ronetwoinv w$, where, given $w\in \HHone$
or $\HHtwo$, we have $U=\pi_0U$ given by \eqref{pidef} with
\be\label{5-rinv}
\pmatrix{\US_1(k)\cr \US_2(k)}= \vonetwo(k)^{-1}\hat{w}(k) = 
\frac{1}{2ik\lambda_1(k)}
\pmatrix{\lambda_1(k)&ik\cr\lambda_1(k)&-ik}
\pmatrix{\hat{w}_1(k)\cr\hat{w}_2(k)}
\ee
for $k\ne0$. Clearly
\[
|\US_1(k)|^2+|\US_2(k)|^2=
|\vonetwo(k)^{-1}\hat{w}(k)|^2 \le \frac{C}{1+k^2}|\hat{w}(k)|^2
\]
for all $k$, so using \eqref{normeq} we see that
$\Ronetwoinv$ is bounded from $\HHone$ to $H_0$ and from
$\HHtwo$ to $X_0$.
\qed

\noindent
{\bf Proof of Theorem \ref{th:init}.}
Given $w$, we shall prove corresponding statements for 
$\zeta=\Ronetwo\xi$ in place of $\xi$ and use Lemma~\ref{lm:r12}.
Recall that $\Ronetwoinv$, the inverse of $\Ronetwo$, is given by
$U=\Ronetwoinv w$ using \eqref{5-rinv} and \eqref{pidef}.
The quantity $\zeta$ should satisfy
\be\label{5-zeta}
\zeta = w-\ponetwo(\zeta)\qquad\hbox{where}\quad
\ponetwo(\zeta)=\Ronetwo \phd(\Ronetwoinv\zeta).
\ee
Using the contraction mapping theorem, we find that this
equation has a unique solution $\zeta\in\HHone$ satisfying
$\|\zeta\|_\HHone<\delta'$ if $\|w\|_\HHone<\delta'(1-L(\delta'))$,
provided that the Lipschitz constant $L(\delta')$ of $\ponetwo$
on the ball of radius $\delta'$ in $\HHone$ satisfies $L(\delta')<1$.
This is true if $\delta'$ is sufficiently small, due to Theorem~\ref{T:1}(ii)
and Lemma~\ref{lm:r12}. 
Moreover, the map $w\mapsto\zeta$ is Lipschitz with Lipschitz constant 
$1/(1-L(\delta'))$. Furthermore, $\ponetwo(\zeta)\in\Ronetwo
X_1\subset \HHtwo$, so if $w\in\HHtwo$ then $\zeta\in\HHtwo$ and so
$\xi=\Ronetwoinv\zeta\in X_0$. 
\qed

\noindent
{\bf Proof of Theorem~\ref{th:main}.}
The parts of the Theorem referring to existence, uniqueness,
stability, and Lipschitz dependence on initial data follow directly from 
Theorems~\ref{th:stab} and \ref{th:init}. It remains to verify that
the first two components $w=\Ronetwo U$ of a solution as given by
these results satisfy an equation of the form \eqref{eq:wave}. 
Let $\xi(y)$ be determined from $w(y)$ by Theorem~\ref{th:init} for
each $y$. Since $\Ronetwoinv\Ronetwo\xi=\xi$, we find that
\be\label{eq:uw}
U = \Ronetwoinv w +\wtoU(w) \qquad\hbox{where}\quad
\wtoU(w) = (I-\Ronetwoinv\Ronetwo)\phd(\xi).
\ee
Note that $\phd$ takes values in $X_1$ so $\wtoU(w)$ need not
be zero.  However the first two components $\Ronetwo\wtoU(w)=0$,
and since $U$ is a classical solution of \eqref{eq:stw}, restriction
to the first two components yields 
\be\label{eq:w1}
\frac{d}{dy} w= \Ronetwo A (\Ronetwoinv w+\psi(w)) = \Aonetwo w + 
\pmatrix{ 0\cr \psi_3(w)}
\ee
since the first two components $\Ronetwo f(U)=0$. 
Here the action of the operator $\Aonetwo=\Ronetwo A\Ronetwoinv$
can be determined through Fourier transform representation from
\eqref{3-urep} and \eqref{5-rinv}. We find that
\begin{eqnarray}
\Aonetwo w &=& \sum_{k\in\Z} e^{ikx} \vonetwo(k)
\pmatrix{\lambda_1(k)&0\cr0&-\lambda_1(k)} \vonetwo(k)^{-1} \hat w(k)
\nonumber \\
&=& \sum_{k\ne0} e^{ikx}
\pmatrix{0 & ik\cr \lambda_1(k)^2/ik &0} 
\pmatrix{\hat w_1(k)\cr \hat w_2(k)}.
\label{5-Arep}
\end{eqnarray}
Thus we have that $w=\Ronetwo U$ satisfies an equation of the form
\eqref{eq:wave} in which $g=\wtoU_3$ and 
$S$ is a pseudodifferential operator of degree zero defined by
\be
\label{eq:Shat}
\widehat{Sw_1}(k)=
(\lambda_1(k)/ik)^2 \hat w_1(k)
\ee
for $k\ne0$.
The eigenvalues of $\Aonetwo$ have the form $\pm\lambda_1(k)$,
leading to the real dispersion relation in \eqref{1-lsq} 
with the plus sign.
Since $\phd$ takes values in $X_1$ and satisfies $\phd(0)=0$,
$D\phd(0)=0$, we find that $g(w_1,w_2)=\psi_3(w)$ is Lipschitz from 
a small ball in $\HHone$ into $\IH_1$ with $g(0,0)=0$, $Dg(0,0)=0$.

This finishes the proof of Theorem \ref{th:main}. \qed

\setcounter{equation}{0}

\appendix

\section{\bf Center manifolds of infinite dimension and codimension}

Here we consider abstract differential equations of the form
\begin{equation} \label{eq:FODS}
\frac{du}{dt}(t) = Au(t) + f(u(t)).
\end{equation}
where $X$ and $H$ are Banach spaces with $X$ densely embedded in $H$, 
$A\in \IL(X,H)$, the space of bounded linear operators from
$X$ to $H$, and $f$ is continuously differentiable 
from $H$ into $X$ with $f(0)=0$ and $Df(0)=0$.

The goal in this section is to prove the existence of a locally
invariant center manifold of classical solutions 
for the system \eqref{eq:FODS} under certain
conditions which permit the center subspace (that associated with the
purely imaginary spectrum of $A$) to have infinite dimension and
codimension. 

We start with some basic definitions and some hypotheses:

\begin{Def} Let $J \subset \R$ be an open interval and  
$ u\colon {{\R}} \to H $ be a function. We say that $u$ is a classical 
solution of \eqref{eq:FODS} on $J$ if the mapping $t \mapsto u(t)$ is
continuous from $ J$ into $ X$, is differentiable from $J$ into $H$
and \eqref{eq:FODS} holds for all $t\in J$.
\end{Def}
Let $\beta>0$, let  $Y$ and $Z$ be Banach spaces and $U$ be an open set in $Y$. 
We define the Banach spaces $C_{b}(U,Z)$, $\Lip(U,Z)$ and $Y^\beta$ by
\begin{eqnarray*}
C_{b}(U,Z) &:=& \left\{ f \in C(U,Z): \sup_{u\in Z}\|f(u)\|_{Z} < \infty  \right\}. \\
\Lip(U,Z) &:=& \left\{ f \in C(U,Z): \|f(u)-f(v)\|_{Z} \leq M_f \|u - v\|_{Y}
\ \mbox{for all} \ u, v \in U \right\}.\\
Y^{\beta} &:=& \{u \in C({{\R}},Y): ||u||_{Y^\beta}:=\sup_{t}
e^{- \beta |t|} \|u(t)\|_{Y} < \infty \} .
\end{eqnarray*}

Throughout this section we assume that there are bounded
projections $\pi_{0}$ and $\pi_{1}$ on $H$ such that
(i) $H = H_0 \oplus H_1$ with $H_i := \pi_i(H)$,
(ii) $\pi_{i}|_{X}$ is bounded from $X$ to $X$, and
(iii) $AX_i \subseteq H_i$ where $X_i := \pi_i(X) $, for $i=0, 1$.
We let $\pibd$ denote a common norm bound for the projections $\pi_j$
on $H$ and $X$, $j=0, 1$.

In consequence, the equation \eqref{eq:FODS} can be rewritten as the 
first order system
\be\label{eq:Asys}
\begin{array}{rll}
\displaystyle{\frac{d}{dt}}u_0(t) & = & A_0u_0(t) + \pi_0 f(u(t)), \\[8pt] 
\displaystyle{\frac{d}{dt}}u_1(t) & = & A_1u_1(t) + \pi_1 f(u(t)),
\end{array}
\ee
where  $A_i \in \IL(X_i, H_i)$ with $A_i y = \pi_i Ay$ for $y \in X_i$.

We assume the following splitting properties for the operator $A$,
associated with the linear evolution equation $du/dt=Au$.

\begin{itemize}
\item[(H0)] 
$A_0$ is the generator of a $C^0$-group $\{\Sc(t)\}_{t \in {\R}}$ on 
$H_0$ with subexponential growth. I.e., 
given any $\beta >0$, there is a constant $\Mc >0$ such that 
\[
 \|\Sc(t)\|_{\IL(H_0)} \leq \Mc  e^{ \beta|t|} \quad \Forall t\in {\R}.
\]

\item[(H1)] There exists $\alpha>0$ and a positive
function $M_1$ on $[0,\alpha)$ such that for any $\beta\in[0,\alpha)$
and for any $g_1\in C(\R,X_1)\cap H_1^\beta$ the equation 
\be\label{eq:hyp}
\frac{d}{dt}u_1 = A_1u_1+g_1
\ee
has a unique solution in $H_1^\beta$ given by $u_1=K_1g_1$, where
$K_1\in \IL(H_1^\beta)$ with $\|K_1\|_{\IL(H_1^\beta)}\le \Mh$. 
Furthermore $\|K_1\|_{\IL(X_1^\beta)}\le \Mh$.
\end{itemize}

\begin{Thm}[Local Center Manifold Theorem] \label{T:1}
Let $H$, $ X$, $A$, $\pi_0$, $\pi_1$ and $f$ be as above,
and let
\[
B(\delta) =\{y\in H_0 : \|y\|_{H}< \delta \}.
\] 
Then for all sufficiently small $\delta >0$ there exists 
$\phd \colon H_0\to X_1$ such that
\begin{itemize}

\item[(i)] $\phd(0)=0$ and $D\phd(0)=0$.

\item[(ii)] 
$\phd\in C_b(H_0,X_1)\cap \Lip(H_0, X_1)$, and on any ball
$B(\delta')$, $\phd$
has Lipschitz constant $L(\delta')$ satisfying $L(\delta')<\frac12$  
and $L(\delta')\to0$ as $\delta'\to0^+$.

\item[(iii)] The manifold $\md \subset X$ 
given by 
\be\label{eq:md}
\md := \{\xi + \phd(\xi): \xi\in X_0 \}
\ee
is a local integral manifold for \eqref{eq:FODS} over 
$B(\delta)\cap X_0$.
That is, given any  
$y\in \md$ there is a continuous map $u \colon \R \to \md$ with
$u(0)=y$, such that for any open interval $J$ containing $0$ with 
$\pi_0 u(J) \subset B(\delta)$ it follows that 
$u$ is a classical solution of \eqref{eq:FODS} on $J$.
Moreover, $u_0:=\pi_0u$ is the unique classical solution on $J$
with $u_0(0)=\pi_0y$ to the reduced equation
\be\label{eq:u0}
 \frac{d}{dt}u_0(t) = A_0u_0(t) + \Fd(u_0(t)),
\ee
where $\Fd\colon H_0\to X_0$ is locally Lipschitz and
is given by $\Fd(w):=\pi_0 f(w+\phd(w))$.

\item[(iv)] For any open interval $J\subset\R$,
every classical solution $u_0\in C^1(J,H_0)\cap C(J,X_0)$
of the reduced equation \eqref{eq:u0}
such that $u_0(t) \in B(\delta)$ for all $t \in J$
yields, via $u=u_0 + \phd(u_0)$,  a classical solution $u$ 
of the full equation \eqref{eq:FODS} on $J$.

\item[(v)] The manifold $\md$ contains all classical solutions on $\R$
that satisfy 
$\|u(t)\|_{H} \leq \delta$ for all $t$.
\end{itemize}
\end{Thm}

Center manifolds of infinite dimension with finite codimension
(dim $H_1 <\infty$) were obtained by Bates and Jones \cite{BJ}.
A generalization regarding invariant manifolds of
infinite dimension and codimension in nonautonomous systems
was obtained by Scarpellini \cite{Sc90},
but his hypotheses require that the operator $A_1$ be bounded from $H$ to $H$.
The general strategy of our proof will follow closely the lines of
\cite{VI89} (also see \cite{Ki82,Mi88a,Mi92})
for the case of a finite-dimensional center manifold
in an ill-posed system for which the spectrum of $A_1$ is unbounded
on both sides of the imaginary axis.
One transforms Eq.~\eqref{eq:FODS} into an integral equation
that must contain all small bounded solutions. In order to 
obtain an invariant manifold by a contraction mapping argument,
one must modify the nonlinearity $f$ outside a neighborhood of $0$
using a cutoff function.

A significant point of difference between our results and those
of \cite{VI89} is that our cutoff occurs in the $H$ norm, and 
not in the $X$ norm as in \cite{VI89}.
In our application to traveling waves of the Benney-Luke equation,
it is important to establish global existence of small solutions
on the center manifold. We do this by using an energy
functional which is defined on $H$ and conserved in time for 
classical solutions (which take values in $X$). 
The energy is indefinite in general, but controls the $H$ norm
for solutions on the center manifold.
For this reason, we find it necessary to obtain a center manifold
that contains solutions with large $X$ norm but small $H$ norm.
This we accomplish by requiring that the nonlinearity $f$ has
a smoothing property, mapping $H$ into $X$. This is a stronger
hypothesis on $f$ than is made in earlier works, but 
we want to emphasize that it is completely natural for the 
present application to the Benney-Luke equation.

\subsection{\bf Linear Analysis}

In this section we discuss the existence of classical
solutions for inhomogeneous linear problems 
that are associated with the system \eqref{eq:Asys}.
First we begin with the linear analysis corresponding to the problem on the 
center space. 
The following result follows easily from the
standard theory of $C^0$ semigroups (see \cite{Pa}).

\begin{Lem} (Linear analysis on the center space) \label{L.center}
Suppose condition (H0) holds. 
Then for any $ \zeta \in H_0$ and $g_0\in H_0^{\beta}$ the initial value problem
\be\label{eq:III}
{\frac{d}{dt}}u_0(t) = A_0u_0(t) + g_0(t), \quad u(0) = \zeta, 
\ee
has a global mild solution $u_0\in H_0^{\beta}$ given by
\begin{equation} \label{eq:form}
u_0(t) = \Sc(t)\zeta + \int^{t}_{0} \Sc(t-\tau) g_0(\tau)d \tau .
\end{equation}
Moreover, if  $g_0\in C(\R,X_0)$ 
then for every $\zeta \in X_0$ 
Eq.~\eqref{eq:III} has a unique global classical solution.
In either case $Y=H_0$ or $X_0$, for $g_0\in Y_\beta$ we have the 
estimate
\begin{equation}\label{eq:Cest}
||u_0||_{Y^\beta}  \leq \Mc \|\zeta\|_{Y} 
+\frac{M_0(\beta/2) }{\beta/2} ||g_0|| _{Y^\beta}.  
\end{equation}
\end{Lem}

Now we can treat the full linear problem on $H$
by simply combining the results obtained in the center space
and the hypothesis (H1) in the hyperbolic space. 

\begin{Lem}(Combined linear analysis) \label{L.combined}
Let $H, X, A, \pi_{0}, \pi_1$ be as above and 
let $\beta \in (0, \alpha)$ be fixed.
Then for every $\zeta \in X_0$ and for every 
$g \in H^{\beta} \cap C(\R,X)$, the problem
\be\label{eq:Lin}
\frac{d}{dt}u(t)  =  Au(t) + g(t) \quad (t \in {\R}), 
\qquad
\pi_0u(0)  =  \zeta
\ee
has a unique classical solution $u \in H^{\beta}$ given by
\begin{equation} \label{eq:mild}
u(t) = \Sc(t) \zeta + \int_{0}^{t} \Sc(t-\tau)\pi_0g(\tau) d \tau +
(K_1\pi_1g)(t).
\end{equation}
Moreover, 
if $g \in X^{\beta}$, the $u$ given by \eqref{eq:mild} is in $X^{\beta}$.
In either case $Y=X$ or $H$ we have the estimate 
\be\label{eq:ubd1}
||u||_{ Y^\beta} \leq \Mc  \| \zeta \|_{Y} +
\Mt ||g||_{ Y^\beta}
\ee
where 
\[
\Mt:= \pibd\left(
\frac{M_0(\beta/2) }{\beta/2} + \Mh \right).
\]
\end{Lem}

Due to hypothesis (H1) and Lemma~\ref{L.center}, we have that
in fact for any $\zeta\in H_0$ and $g\in H^\beta$, 
formula \eqref{eq:mild} defines a function $u\in H^\beta$
that satisfies the bound in \eqref{eq:ubd1}. We will call
this function $u$ the {\it mild} solution of \eqref{eq:Lin}.

\subsection{\bf Nonlinear analysis with cutoff}

In this section, we will consider the full nonlinear problem. 
The first observation is that by Lemma \ref{L.combined}, any
classical solution 
$u$ of \eqref{eq:FODS} that is globally bounded in $H$ must 
satisfy the equation
\be\label{eq:fint}
u(t) = \Sc(t) \zeta + 
\int_{0}^{t} \Sc(t-\tau)\pi_0f(u(\tau)) \,d\tau
+(K_1\pi_1 f(u))(t),
\ee
where  $\zeta= \pi_0 u(0)\in H_0$. The idea now is to use
the contraction mapping theorem in the space $H^{\beta}$ to prove the
existence of a unique fixed point for the operator that yields the
right hand side of Eq.~\eqref{eq:fint}.
But the nonlinearity $f$ may be only locally, not globally, 
Lipschitz. This forces us to localize Eq.~\eqref{eq:FODS} by changing the
nonlinearity outside of a suitable neighborhood of $0$ in the space $H$ 
using a cutoff function. 
We will establish a center manifold theorem for
the cutoff version of Eq.~\eqref{eq:FODS}, 
and then show that this produces a local center manifold for the full problem. 

Recall $f \in C^{1}(H,X)$  satisfies  $f(0)=0$, $Df(0)=0.$ 
Choose $ \chi \in C^{1}(\R,[0,1])$ such that 
\be\label{d:chi}
\chi(t)=\cases{1& for $t<2$,\cr 0& for $t>3$,}
\ee
with $|\chi'(t)|\le 2$ for all $t$.
For $\delta >0$, then define the cut-off nonlinearity by
\begin{equation} \label{loc-f}
\fdel(x) := f(x) \cdot \chi\left(\frac{\|x\|_{H}}{\delta} \right)
\end{equation}
for $x \in H$. 
Then $\fdel$ is bounded, globally Lipschitz, and is $C^1$ near zero, with
$\fdel(x)=0$ for $\|x\|_{H} > 3\delta$ and $D\fdel(0)=0$. 
We next define 
\begin{eqnarray}
L(f,\delta) &:= &
\sup\{\|Df(y)\|_{\IL(H,X)}: y\in H, \|y\|_H<\delta\},
\label{D.lip}\\[4pt]
L(\fdel) &:=&
\sup\{\|\fdel(x)-\fdel(y)\|_X/\|x-y\|_H : x,y\in H\ \mbox{with $x\ne y$}\}.
\end{eqnarray}
For any $\delta'>0$ we have
\be\label{f.lip}
\|f(u_1)-f(u_2)\|_{X} \le L(f,\delta')\|u_1-u_2\|_H
\quad\mbox{for all $u_1, u_2\in B(\delta')$,}
\ee
and since $Df \in C(H, \IL(X))$ 
we have $L(f,\delta')\to0$ as $\delta'\to0^+$. 
Our cutoff yields the bounds
\be\label{fd.lip}
L(\fdel) \le 7L(f,3\delta), \qquad
L(\fdel,\delta') \le 3L(f,\delta')
\quad\mbox{for $0<\delta'\le\delta$.}
\ee

The cut-off version of Eq.~\eqref{eq:FODS} is
\begin{equation} \label{eq:fdel}
\frac{du}{dt}(t) = Au(t) + \fdel(u(t)).
\end{equation}
By Lemma \ref{L.combined}, any classical solution  
$u \in H^{\beta}$ of \eqref{eq:fdel} 
must satisfy the equation 
\begin{equation}
u(t) = \fix(\zeta, u)(t) : = \Sc(t) \zeta + \tfix(u)(t) \label{eq:lin} 
\end{equation}
where
\[
\tfix(u)(t)  := \int_{0}^{t} \Sc(t-\tau)\pi_0\fdel(u(\tau)) d \tau
+(K_1\pi_1 \fdel(u))(t),
\]
and where $ \zeta=\pi_0u(0)\in H_0$.

By Lemma~\ref{L.combined},
if $\beta \in (0,\alpha)$ then 
$\tfix\colon H^{ \beta} \to X^{ \beta} $ 
is Lipschitz continuous with Lipschitz constant 
$\Mt L(\fdel)$,
and in either case $Y=H$ or $X$ we have that
$\fix(\zeta,\cdot)\colon Y^{ \beta} \to Y^{ \beta} $
is Lipschitz continuous with the same Lipschitz constant.
Let $\delta_{0}$ be chosen (depending on $\beta$) 
such that for $ 0 < \delta \le \delta_{0} $,
\begin{equation}\label{delta1}
\Mt L(\fdel)
<\sfrac12.
\end{equation}
Now we can solve Eq.~\eqref{eq:lin} 
by the contraction mapping theorem in $Y^{\beta}$, and obtain

\begin{Prop}(Fixed points of $\fix$) \label{C:4}
Let $ H, X, A,  \pi_{0}, \pi_1 , f$ and $\delta$  be as
above. 
If  $\beta \in (0,\alpha)$ 
is fixed, then the cut-off nonlinear problem \eqref{eq:fdel}
has for all $\zeta \in H_{0} $ a unique mild solution 
$u=: \umap(\zeta,\cdot) \in H^{\beta}$ of the form \eqref{eq:lin} with
$\pi_{0}u(0) = \zeta $. 
If $\zeta \in X_0$ then $\umap(\zeta,\cdot) \in X^{\beta}$ and is the unique
classical solution in $H^{\beta}$.
Moreover, in either case $Y=H$ or $X$,
for any $\zeta_1, \zeta_2 \in Y_{0}$ we have 
\begin{equation}
||\umap(\zeta_1,\cdot)- \umap(\zeta_2,\cdot)||_{Y^\beta}  \leq
2\Mc \| \zeta_1 - \zeta_2 \|_{Y}.
\end{equation}
\end{Prop}
{\bf Proof.}
We only need to show the estimates (the existence and 
uniqueness follow by the contraction mapping theorem).
By estimates in Lemma \ref{L.combined}, 
\begin{eqnarray*} 
||\umap(\zeta,\cdot) -\umap(\eta,\cdot) ||_{Y^\beta}  &\leq&
\Mc \| \zeta - \eta \|_{Y} +
\Mt
||\fdel(\umap(\zeta,\cdot)) - 
\fdel(\umap(\eta,\cdot))||_{Y^\beta}
\\
&\leq&
\Mc \| \zeta - \eta \|_{Y} +
\Mt L(\fdel)||\umap(\zeta,\cdot) -\umap(\eta,\cdot) ||_{Y^\beta}. 
\end{eqnarray*}
\qed

Now we can obtain a global center manifold for the cut-off problem.

\begin{Prop}(Global center manifold) \label{C:5}
Let $H$, $X$, $A$, $\pi_0, \pi_1$, $\beta$, $ \delta$ and $f$ be as in 
Proposition \ref{C:4}.
For $\zeta\in H_0$, let $\umap(\zeta,\cdot)$ denote the 
unique fixed point for $\fix(\zeta,\cdot)$ in $H^\beta$,
and define
\[
\psd(\zeta) = \pi_1 \umap(\zeta,0).
\]
If $u \in H^{ \beta} $ is a mild solution of Eq.~\eqref{eq:fdel},
then $u(t)\in \CM$ for all $t \in \R$, where
\[
\CM:= \{ \xi + \psd( \xi): \xi \in H_0 \} .
\]
The map
$\psd \in C_{b}(H_{0},X_{1})\cap \Lip(H_0,X_1)$, 
$ \psd(0)=0$, and $D\psd(0)$ exists and is zero. 
Also, for any $\delta'>0$,
for any  $\zeta_1, \zeta_2 \in H$ with $\|\zeta_j\|_H<\delta'$ we have
\[
\| \psd(\zeta_1) - \psd(\zeta_2) \|_{X} \leq 
L(\psd,\delta')
\|\zeta_1 - \zeta_2 \|_{H} 
\]
where $L(\psd,\delta')\to0$ as $\delta'\to0^+$ and with  
$C_*=2\Mc\Mh\pibd $ we have
$L(\psd,\delta')\le C_* L(\fdel)$ for all $\delta'>0$. 
\end{Prop}

{\bf Proof.}
Let $\zeta \in H_0$, 
then the equation
\[
u = \Sc(\cdot) \zeta + \tfix(u) 
\]
has the unique solution $u=\umap(\zeta,\cdot)$ in $H^{ \beta}$.
Since $\pi_0 \tfix(u)(0) = 0$, we conclude that $\pi_0 u(0) = \zeta $.
This means that if $u\in H^\beta$ is a mild solution of \eqref{eq:fdel} then
$u= \umap(\zeta,\cdot)$ and 
\[
\pi_1u(0)= \pi_1\umap(\zeta,0)= \psd(\zeta).
\]

In particular note the following. Fix any $s\in\R$.
Then in hypothesis (H1), replacing $g_1$ by $g_1(s+\cdot)$ 
yields the solution $u_1(s+\cdot)$.
Now, given $u=\umap(\zeta,\cdot)$ as above, define $v_{s}(t)=u(s+t)$. 
Since $\pi_1u$ is a solution
of \eqref{eq:hyp} with $g_1(t)=\pi_1\fdel(u(t))$, we have that
$\pi_1v_s$ is a solution of
\eqref{eq:hyp} with $g_1(t)=\pi_1 \fdel(v_s(t))$. 
Since $v_{s} \in H^{ \beta}$, 
$v_{s}$ is a mild solution of \eqref{eq:fdel}  with 
$\pi_0v_{s}(0)=\pi_0 u(s)$. 
It follows that $v_{s} = \umap(\pi_0u(s),\cdot)$ and therefore
\[
u(s)=v_s(0)=\pi_0u(s)+\psd(\pi_0u(s)).
\]
This proves that $u(t) \in \CM$ for all $t \in \R$.

To prove the estimate we let $\delta'>0$ and 
suppose $\zeta_1$, $\zeta_2\in B(\delta')$.
Note that we have 
$\psd(\zeta_j)=K_1\pi_1\fdel(\umap(\zeta_j,\cdot))(0)$
and $\|\umap(\zeta_j,t)\|_H\le 2\Mc e^{\beta|t|}\delta'$.
Taking any $\gamma\in[\beta,\alpha)$, we compute
\begin{eqnarray*} 
\|\psd(\zeta_1) - \psd(\zeta_2)\|_{X} 
&\leq& M_1(\gamma) \pibd 
\|\fdel(\umap(\zeta_1,\cdot))
-\fdel(\umap(\zeta_2,\cdot))\|_{X^{\gamma}}
\\[4pt]
&\leq& 
M_1(\gamma) \pibd 
\sup_{t}\left( e^{-\gamma|t|}L(\fdel,2\Mc e^{\beta|t|}\delta')
\|\umap(\zeta_1,t) -\umap(\zeta_2,t)\|_H
\right)
\\
&\leq&
2\Mc M_1(\gamma) \pibd 
\sup_t e^{-(\gamma-\beta)|t|} L(\fdel,2\Mc e^{\beta|t|}\delta')
\|\zeta_1-\zeta_2\|_H
\end{eqnarray*}
Since $L(\fdel,2\Mc e^{\beta|t|}\delta')$ is uniformly bounded in $t$
by $L(\fdel)$ and it approaches zero as $\delta'$ does for $t$ fixed, 
the asserted estimates follow,
and $D\phd(0)=0$. Clearly $\phd(0)=0$.
We also have that $\phd$ is globally bounded, since
\be\label{phdbd}
\|\phd(\zeta)\|_X \le M_1(\gamma) \pibd 
\sup_{u\in H}\|\fdel(u)\|_X \le 3\delta L(\fdel) M_1(\gamma) \pibd .
\ee
\qed

On the center manifold, the evolution reduces to a well-posed
problem for a semilinear problem whose linear part is solved
by the $C^0$-semigroup $\Sc$.

\begin{Prop}(Equation of evolution on the center manifold)
\label{R:2.4}
Make the same hypotheses as in Proposition~\ref{C:5}.
Let $\Fdd(w):=\pi_0 \fdel(w+\phd(w))$ for $w\in H_0$.
Then for any $\zeta\in H_0$,
$u_0= \pi_0 \umap(\zeta, \cdot) \in H^{\beta}_0$ is the unique mild 
solution of 
\be\label{eq:redd}
\frac{d}{dt} u_0(t)  = A_0u_0(t) + \Fdd(u_0(t)) 
\quad (t \in {\R}),\qquad u_0(0) = \zeta .
\ee
If $\zeta \in X_0$, then $u_0= \pi_0\umap(\zeta, \cdot) \in X_0^{ \beta}$
is the unique classical solution for this problem.
For any $T>0$, the map $\zeta\mapsto u_0$ is Lipschitz continuous
from $H_0$ to $C([-T,T],H)$.
\end{Prop}

{\bf Proof.}
Let $\zeta \in H_0$. 
Since $\Fdd$ is Lipschitz on $H_0$ and $A_0$ is the generator
of a $C^0$ semigroup on $H_0$, existence and uniqueness 
of a global mild solution of \eqref{eq:redd}
is proved in a standard way using semigroup theory \cite{Pa}.
By Proposition \ref{C:4},  $\umap(\zeta,\cdot) \in H^{\beta}$ 
is the unique mild solution of \eqref{eq:fdel}
of the form \eqref{eq:lin} with
$\pi_0\umap(\zeta,0)=\zeta$. Now by Proposition \ref{C:5},
\[
\umap(\zeta, t)= \pi_0\umap(\zeta, t) + \psd(\pi_0\umap(\zeta,t)).
\]
Then by projecting \eqref{eq:lin} on the space $H_0$ we see that
$\pi_0\umap(\zeta, \cdot) \in H_0^{\beta}$ is the unique mild solution of 
\eqref{eq:redd}.
If $\zeta \in X_0$ we have that $\pi_0\umap(\zeta, \cdot) \in
X_0^{\beta}$ is the unique classical solution of \eqref{eq:redd}.
That the map $\zeta\mapsto u_0$ is Lipschitz continuous from $H_0$
to $C([-T,T],H)$ for any $T>0$ follows in a standard way from the 
variation of parameters formulation of \eqref{eq:redd}.
\qed

\subsection{\bf Proof of Theorem~\ref{T:1}.}
Let $\delta$ be so small that both 
$\Mt L(\fdel)<\sfrac12$ and $C_*L(\fdel)<\sfrac12$. 
We take $\phd$ as given by Proposition~\ref{C:5}.
Then 
$\phd \in C_{b}(H_0,X_{1})\cap \Lip(H_0,X_1)$.  
Let $\md \subset X $ be defined by
\[ 
\md := \{ \xi + \phd( \xi) : \xi \in X_0 \}.
\]
The statements in parts (i) and (ii) of the Theorem
follow from Proposition \ref{C:5}.

(iii) Let $ y = \zeta + \phd( \zeta) \in \md $. Then there exists a 
unique classical solution $u= \umap( \zeta,\cdot) \in H^{ \beta}$ of 
the cut-off Eq.~\eqref{eq:fdel} with 
$ \pi_{0}\umap( \zeta, 0)= \zeta$. In consequence,
$ \phd( \zeta) = \pi_{1}\umap( \zeta,0)$ and then $ \umap( \zeta, 0)
= \zeta + \phd( \zeta) = y $. Moreover, $u(t) \in \CM$ for all $t \in \R$. 
In other words, $ u(t)= \pi_0u(t) + \psd(\pi_0u(t))$. Let $J$ be any 
open interval containing $0$ with $\pi_0(u(J)) \subset B(\delta)$. Then 
for any $t \in J$,
\[
\|u(t)\|_{H} \leq \|\pi_0u(t)\|_{H} + \|\phd(\pi_0u(t))\|_{H} \leq
             (1 + L(\phd))\|\pi_0u(t)\|_{H} \leq 2 \delta,
\]
where $L(\phd)$ denotes the Lipschitz constant of $\phd$. 
Thus $\fdel(u(t))=f(u(t))$ for all $t \in J$. 
So $u$ is a classical solution of 
the original Eq.~\eqref{eq:FODS} in $J$.

(iv) Let $ u_{0}$ be a classical solution on $\R$ of
\[
\frac{d}{dt} u_{0}(t) = A_0 u_0(t) + \pi_0f(u_0(t) + \phd(u_0(t))),
\]
such that $u_0(t) \in B(\delta) \cap X_0$ for all $t \in {\R}$.
Let $ \zeta= u_0(0) \in X_0 $. Define $w$ by $ w= u_0 + \phd (u_0) $.
We want to show that $w$ is a 
classical solution of the full problem \eqref{eq:FODS}. Let 
$ \umap(\zeta,\cdot) \in H^{ \beta}$ be 
the unique classical solution of the cut-off Eq.~\eqref{eq:fdel}. 
Then by Proposition \ref{C:5}, 
$ \umap( \zeta,\cdot)$ is the fixed point of $\fix(\zeta,\cdot)$ and 
$ \pi_1 \umap(\zeta,t)
= \phd(\pi_0 \umap(\zeta,t))$. Moreover, $ \pi_0
\umap(\zeta,\cdot)$ 
satisfies the above problem. By Proposition \ref{R:2.4}
we conclude that $ \pi_0 \umap( \zeta,\cdot) = u_0$, and hence
$\umap(\zeta, \cdot) = w(\cdot) $. On the other hand,
\[
\|w(t)\|_{H} \leq \|u_0(t)\|_{H}+\|\phd(u_0(t))\|_{H} 
\leq(1+L(\phd))\delta<2\delta.
\]
This implies that $\fdel(w(t))=f(w(t))$ for all $t \in \R$ and $w$ 
is a classical solution of the full problem \eqref{eq:FODS} on $\R$.

(v)  Let $u$ be a classical solution for the full problem \eqref{eq:FODS}
such that $ \|u(t)\|_H \le \delta$ for all $t \in {\R}$. Then
$u$ is also a solution of the cut-off Eq.~\eqref{eq:fdel} in $H^\beta$.
Define $u_0$ as 
$u_0(t) = \pi_0u(t)$. Then by Proposition \ref{C:5} we conclude that
$ u(t) = u_0(t) + \psd(u_0(t))$. Since $u(t)\in X$ we conclude that
$u(t)\in\md$ for all $t$.
\qed

\section*{Acknowledgments}
This material is based upon work supported by the 
National Science Foundation under Grant 
Nos.\  DMS97-04924 and DMS00-72609.
The work of JRQ is supported by the
Universidad del Valle, Cali, Colombia, and
partially supported by Colciencias under Project No.\ 1106-05-10097.


\bibliographystyle{plain}
\bibliography{qptwo0}
\end{document}